\documentclass[12pt, preprint,aps,tightenlines]{revtex4}
\usepackage{epsfig}

\def\beq{\begin{equation}}
\def\eeq{\end{equation}}
\def\bea{\begin{eqnarray}}
\def\eea{\end{eqnarray}} 
\def\beqa{\begin{equation}\begin{array}{l}}
\def\eeqa{\end{array}\end{equation}}

\def\barr{\left(\begin{array}{c}}
\def\earr{\end{array}\right)}
\def\bmat{\left(\begin{array}{cc}}
\def\emat{\end{array}\right)}
\def\al{\alpha}
\def\be{\beta}
\def\ga{\gamma}

\def\De{\Delta}

\def\kp{\kappa}
\def\la{\lambda}

\def\si{\sigma} 
\def\Si{\Sigma}

\def\w{\omega}

\def\bra{\langle} 
\def\ket{\rangle}

\def\mathscr{\mathcal}

\def\vk{\hat{k}}
\def\vkp{{\hat{k}\, '}}

\def\3d{3-D}

\newcommand{\lsim}{\, \, \raisebox{-0.8ex}{$\stackrel{\textstyle <}{\sim}$ }}
 
\newcommand{\lsc}{\Lambda _\chi}
\newcommand{\mpi}{m _\pi}
\newcommand{\cpt}{\chi PT}
\newcommand{\sss}{\scriptscriptstyle }

\newcommand{\vsigone}{{\vec\sigma^{\sss 1}}}
\newcommand{\vsigtwo}{{\vec\sigma^{\sss 2}}}
\newcommand{\veps}{{\hat\epsilon}}
\newcommand{\vepsprime}{{\hat\epsilon\, '}}
\newcommand{\vkay}{{\vec k}}
\newcommand{\vkayprime}{{{\vec k}\, '}}
\newcommand{\vpee}{{\vec p}}
\newcommand{\vpeeprime}{{{\vec p}\, '}}
\newcommand{\vsigma}{{\vec\sigma}}

\newcommand{\barf}{\Upsilon}

\begin{document}

\title{Predictions for Polarized-Beam/Vector-Polarized-Target Observables in Elastic Compton Scattering on the Deuteron}

\author{Deepshikha Choudhury}
\email{choudhur@phy.ohiou.edu}
\author{Daniel R. Phillips}
\email{phillips@phy.ohiou.edu}

\affiliation{
Department of Physics and Astronomy, Ohio University,
Athens, OH 45701}
\date{\today}
\vspace{20.0pt}

\begin{center}
\begin{abstract}
Motivated by ongoing developments at HI$\ga$S at TUNL that include increased 
photon flux and the ability to circularly polarize photons, we calculate 
several beam-polarization/target-spin dependent observables for elastic 
Compton scattering on the deuteron. This is done at energies of the order of 
the pion mass within the framework of Heavy Baryon Chiral Perturbation Theory. 
Our calculation is complete to ${\cal O}(Q^3)$ and at this order there are no free 
parameters. Consequently, the results reported here are predictions of the 
theory. We discuss paths that may lead to the extraction of neutron 
polarizabilities. We find that the photon/beam polarization asymmetry is not a 
good observable for the purpose of extracting $\al_n$ and $\be_n$. However, 
one of the double polarization asymmetries, $\Si_x$, shows appreciable 
sensitivity to $\ga_{1n}$ and could be instrumental in pinning down the 
neutron spin polarizabilities.
\end{abstract}
\end{center}

\maketitle
\thispagestyle{empty}

\section{Introduction}

Compton scattering studies on the nucleon and deuteron have shed light on the 
low-energy dynamics of quarks inside a nucleon. An electromagnetic probe by 
nature, Compton scattering captures information about the response of the 
charge and current distributions inside a nucleon to a quasi-static 
electromagnetic field. These responses are quantified in terms of the nucleon 
polarizabilities. 

To lowest order in photon energy, the spin-averaged amplitude for Compton 
scattering on the nucleon is given by the Thomson term
\begin{equation}
{\rm Amp}=-\frac{{\cal Z}^2 e^2}{M}\hat{\epsilon}\cdot\hat{\epsilon}', 
\label{eq:1}
\end{equation}
where ${\cal Z}e,M$ represent the nucleon charge and mass respectively and
$\hat{\epsilon},\hat{\epsilon}'$ specify the polarization
vectors of the initial and final photons respectively.  At the 
next order in photon energy, contributions arise from electric and
magnetic polarizabilities---$\bar{\alpha}_E$ and
$\bar{\beta}_M$---which measure the response of the nucleon to the application 
of quasi-static electric and magnetic fields. The spin-averaged amplitude is 
expressed as:
\begin{equation}
{\rm Amp}=\veps\cdot\vepsprime\left(-\frac{{\cal Z}^2 e^2}{M}
+\omega\omega' \; 4\pi\bar{\alpha}_E \right)
+\veps\times\vk\cdot\vepsprime
\times\vkp\; \omega\omega' 4\pi\bar{\beta}_M+\; {\cal O}(\omega^4) \; .
\label{eq:2}
\end{equation}
Here, $k_\mu=(\omega,\vec{k})$, ${k'}_\mu=(\omega',\vec{k}')$ 
specify the four-momenta of the initial and final photons respectively. The 
associated differential scattering cross section on the proton is given by 
\begin{eqnarray}
{d\sigma\over d\Omega}&=&\left({e^2\over 4\pi M}\right)^2
\left({\omega'\over \omega}\right)^2\left[{1\over 2}
(1+\cos^2\theta)\right.\nonumber\\
&-&\left.{4\pi M\omega\omega'\over e^2}\left({1\over 2}
(\bar{\alpha}_E+\bar{\beta}_M)(1+\cos\theta)^2+{1\over
2}(\bar{\alpha}_E-\bar{\beta}_M)(1-\cos\theta)^2\right)+\ldots\right].
\nonumber\\
\quad
\end{eqnarray}
It is essential to mention here that for the purpose of this paper, 
$\bar{\alpha}_E$ and $\bar{\beta}_M$ are the so-called Compton 
polarizabilities. In the non-relativistic limit, these reduce to to the static 
electromagnetic polarizabilities. From here on we shall drop the subscripts 
$E$ and $M$ as well as the bars over $\al$ and $\be$.

At the next order in $\w$, we can access the spin polarizabilities ($\ga$'s) 
of the nucleon. Unlike $\al$ and $\be$, however, there is no simple classical 
definition for the spin polarizabilities. Ref.~\cite{barry} provides a 
treatise on polarizabilities for introductory purposes.  

Effective Field Theory ($EFT$) techniques have been successfully used to explore 
the infrared regime of QCD. They are rooted in the symmetries of the 
underlying theory. Research programs in the recent past have facilitated the 
development of a framework based on $EFT$ techniques that allows a consistent 
description of low-energy hadronic and nuclear processes. This framework 
involves building an $EFT$ from the most general Lagrangian involving baryons 
and mesons below the chiral symmetry breaking scale, $\lsc$. In its simplest 
form, this $EFT$ involves the lightest mesons and baryons, namely pions and 
nucleons, as the only degrees of freedom. The heavier mesons and baryons are integrated 
out of the theory. A perturbative expansion is then built as a function of the 
ratio of two scales -- one light and the other heavy -- to be identified with 
the degrees of freedom in the theory. The price one has to pay with such a 
choice is that the domain of validity of the theory is limited by the lightest 
mass scale not explicitly included in the theory. In this case, it is the 
energy required to excite the $\De$(1232) isobar, which is of the order of 
300 MeV.  

This $EFT$ is known as Chiral Perturbation Theory ($\cpt$). In this theory pions 
interact via derivative interactions and/or insertions of the quark mass 
matrix. The hierarchy of scales in this theory is set by the `small' momentum, 
$p$, and the pion mass, which explicitly appears in the interactions, $i.e.$ 
$Q=(p,\mpi)$, and $\lsc$. Thus the $S$-matrix elements can be expressed as 
expansions in powers of the ratio of the two scales, $Q=(p,\mpi)\over \lsc$. 
The dynamics at distances shorter than the wavelength of the probe or those 
involving mesons other than the pions are accounted for by local operators. In 
the purest form of $\cpt$ the strengths (in other words, the coefficients) of 
these operators, called ``low-energy constants'' (LECs) should be extracted 
from experiments. The resulting theory can then be used to predict the outcome 
of other experiments. Baryons are usually included in $\cpt$ by treating them 
as heavy objects because of the fact that the momenta involved are very small 
compared to the baryon rest masses. The resulting theory is called Heavy 
Baryon Chiral Perturbation Theory ($HB\cpt$). A detailed discussion can be 
found in Refs.~\cite{wein,bkmrev}. 

$HB\cpt$ has been very successful in the single-nucleon sector and its 
successes include Compton scattering on a proton. Nucleon Compton scattering 
has been studied in $HB\cpt$ in Ref.~\cite{ulf1}, where the following results 
for the polarizabilities were obtained to ${\cal O}(Q^3)$:
\begin{eqnarray}
\alpha_p=\alpha_n=\frac{5 e^2 g_A^2}{384 \pi^2 f_\pi^2 m_\pi} &=&12.2 \times
10^{-4} \, {\rm fm}^3, \label{eq:alphaOQ3} \nonumber \\ 
\beta_p=\beta_n=\frac{e^2 g_A^2}{768 \pi^2 f_\pi^2 m_\pi}&=& 1.2 \times 
10^{-4} \, {\rm fm}^3, \label{eq:betaOQ3} \nonumber \\
\ga_{1p}=\ga_{1n}=\frac{e^2 g_A^2}{98 \pi^3 f_\pi^2 m_\pi^2} &=&4.4 \times
10^{-4} \, {\rm fm}^4, \label{eq:gamma1OQ3} \nonumber \\ 
\ga_{2p}=\ga_{2n}=\frac{e^2 g_A^2}{192 \pi^3 f_\pi^2 m_\pi^2} &=&2.2 \times
10^{-4} \, {\rm fm}^4, \label{eq:gamma2OQ3} \nonumber \\ 
\ga_{3p}=\ga_{3n}=\frac{e^2 g_A^2}{384 \pi^3 f_\pi^2 m_\pi^2} &=&1.1 \times
10^{-4} \, {\rm fm}^4, \label{eq:gamma3OQ3} \nonumber \\ 
\ga_{4p}=\ga_{4n}=-\frac{e^2 g_A^2}{384 \pi^3 f_\pi^2 m_\pi^2} &=&-1.1 \times
10^{-4} \, {\rm fm}^3. \label{eq:gamma4OQ3}
\end{eqnarray}  
Here, we have used $g_A = 1.26$ for the axial coupling of the nucleon, and 
$f_{\pi} = 93$ MeV as the pion decay constant. At this order, the 
polarizabilities are manifestations of isoscalar pion loops (see 
Fig.~\ref{fig3}), and so are equal for the proton and the neutron. Note that 
the polarizabilities are {\it predictions} of $HB\cpt$ at this order.  The ${\cal O}(Q^3)$ values of the polarizabilities are modified by higher-order corrections which include what can be quite sizeable effects due to the $\Delta$ isobar~\cite{hemmert,gellas}. In our calculations, the ${\cal O}(Q^3)$ values of the polarizabilities will serve as a reference point as we seek to understand how such shifts away from these values (Eq.~(\ref{eq:alphaOQ3})) will impact observables.

There has been an avalanche of experimental data on unpolarized Compton 
scattering on the proton in the recent past at photon energies below 200 MeV~\cite{pexp, olmos}. The proton electromagnetic polarizabilities are now quite 
well established. The current Particle Data Group (PDG) values for 
the proton are:
\begin{eqnarray}
\alpha_p&=&(12.0 \pm 0.6) \times 10^{-4} \, {\rm fm}^3, \nonumber \\ 
\beta_p&=& (1.9 \pm 0.5) \times 10^{-4} \, {\rm fm}^3.
\label{eq:pexp}
\end{eqnarray}
Theoretical efforts have matched experiments with calculations being done up 
to the fourth order in $Q$~\cite{judith, silas2}. The $HB\cpt$ calculations 
give a very impressive overall description of the data for $\w, \sqrt{|t|} 
\lsim$ 200 MeV.

In contrast, due to the lack of free neutron targets, information about the 
neutron polarizabilities has to be gathered through alternative methods. One 
such method involves scattering neutrons on lead to access the Coulomb 
field of the target and examining the cross-section as a function of energy. 
Currently there is much controversy over what this technique gives for 
$\al_n$. Two experiments using this same technique obtained different results 
for $\al_n$ (Refs.~\cite{nexp1, nexp2}).
\begin{eqnarray}
\alpha_n&=&(12.6 \pm 1.5 \pm 2.0) \times 10^{-4} \, {\rm fm}^3; 
\label{eq:nexp1} \\ 
\alpha_n&=& (0.6 \pm 5.0) \times 10^{-4} \, {\rm fm}^3.
\label{eq:nexp2}
\end{eqnarray}
Enik et al~\cite{enik} revisited Ref.~\cite{nexp1} and recommended a value 
of $\al_n$ between $7 \times 10^{-4}$ fm$^3$ and $19 \times 10^{-4}$ 
fm$^3$.

Also, quasi-free Compton scattering from the deuteron was measured at Mainz~\cite{kolb} for incident photon energies of $E_{\ga} = 236 - 260$ MeV and 
one-sigma constraints on the polarizabilities were reported to be:
\begin{eqnarray}
\alpha_n&=&(7.6 - 14.0) \times 10^{-4} \, {\rm fm}^3, 
\nonumber \\ 
\beta_n&=& (1.2 - 7.6) \times 10^{-4} \, {\rm fm}^3.
\label{eq:nexp3}
\end{eqnarray}

Note that it is necessary to pin down only one of $\al_n$ and $\be_n$. In 
practice, one uses the dispersion sum rule that relates the sum of the 
polarizabilities to the total photoabsorption cross-section in order to 
extract the other. For the proton~\cite{olmos},
\begin{equation}
\al_p + \be_p = \frac{1}{2 \pi^2} \int_{\w_{th}}^{\infty} \frac{\si_p^{tot} 
(\w)}{\w^2}\, \mathrm{d}\w = (13.8 \pm 0.4) \times 10^{-4} \, {\rm fm}^3.
\label{eq:baldinp}
\end{equation}
where, $\si_p^{tot}(\w)$ is the total photoabsorption cross-section and 
$\w_{th}$ is the pion-production threshold. In Ref.~\cite{babu} the sum rule 
for the neutron was extracted from the deuteron photoabsorption cross-section 
by subtracting the proton contribution obtaining,
\begin{equation}
\al_n + \be_n = (14.4 \pm 0.66) \times 10^{-4} \, {\rm fm}^3.
\label{eq:baldinn}
\end{equation}

Given the discrepancy between the measurements of $\al_n$ in Eqs.~(\ref{eq:nexp1}) and (\ref{eq:nexp2}), it is important to measure it by other 
means. Compton scattering on a nuclear target is an obvious candidate. 
However, processes with $A>1$ pose a different kind of challenge because they 
require an understanding of the inter-nucleon interaction. For example, 
deuteron structure is governed by the $NN$ interaction and hence, when analysing $\ga d$ data, one has to 
include the effects of two-body currents over and above the single-nucleon 
$\gamma N$ amplitude. There are different schools of thought regarding this. 
The first school was that introduced by Weinberg~\cite{wein} and later 
refined. Another formulation where pions are treated perturbatively was 
developed by Kaplan, Savage and Wise (KSW)~\cite{ksw}. However, the KSW 
scheme is known to break down at much lower energies than the Weinberg 
formulation. We use the Weinberg formulation for our calculations. Elastic 
Compton scattering on the deuteron has been studied in this theory up to 
${\cal O}(Q^4)$~\cite{silas2, silas1} as a means to extract the spin-independent 
nucleon isoscalar polarizabilities.

Recent experimental advances have made it possible to venture into the area of 
polarization observables. Technology to polarize targets and the current 
developments at the HI$\ga$S facility at TUNL~\cite{gao} have especially 
motivated research in this area, in particular the research reported in this 
paper. The HI$\ga$S upgrade program includes increased photon flux and the ability 
to circularly polarize photons. We believe that these efforts will make it 
possible to design experiments to access the spin polarizabilities of the 
nucleon. Hildebrandt et al~\cite{robert1} have studied polarization 
observables for Compton scattering on the nucleon (both proton and neutron) 
with a focus on the spin polarizabilities. Their calculation involves free 
nucleons and is beneficial for understanding processes that involve quasi-free 
kinematics like $\ga d \rightarrow \ga np$ as opposed to $\ga d \rightarrow 
\ga d$. In this paper, we calculate {\it elastic} Compton scattering on the 
deuteron to ${\cal O}(Q^3)$ but focus on specific observables so as to 
construct road-maps to extract the neutron polarizabilities, $\al_n$, $\be_n$ 
and $\ga_{1n} \ldots \ga_{4n}$. These calculations are exploratory, and are expected to provide a reference for polarization observables for elastic Compton scattering on the deuteron. As such they will serve to focus further explorations of these observables at ${\cal O}(Q^4)$ and beyond.

Until now, there have been only limited efforts to measure nucleon spin 
polarizabilities. The only measurements attempted have been for the forward 
and backward spin polarizabilities. The experimental values for the backward and the forward spin polarizabilities quoted later in this Section include the contribution from the t-channel $\pi^0$ pole. However, deuteron being an isoscalar, this graph does not contribute in the theory discussed in this paper.

The proton backward spin polarizability, $\ga_{\pi} = \ga_1+\ga_2+2\ga_4$ has 
been extracted by the LEGS collaboration from unpolarized Compton scattering 
on the proton~\cite{tonni}. But, the extracted value
\begin{equation}
\ga_{\pi p} = (-27.1 \pm 2.2) \times  10^{-4} \, {\rm fm}^4,
\label{eq:gpp}
\end{equation}
contradicts predictions of standard dispersion theory~\cite{lvov, drech, 
babu2} and $\cpt$~\cite{hemmert, gellas, kumar}. The theoretical prediction 
for $\ga_{\pi p}$ from $\cpt$ is $-36.7 \times 10^{-4}$ fm$^4$~\cite{hemmert, 
vlad}. The LEGS result is also in disagreement with the TAPS result 
$\ga_{\pi p} = (-35.9 \pm 2.3) \times  10^{-4}$ fm$^4$~\cite{olmos}. The most 
recent Mainz experiment ~\cite{camen} using the Mainz 48 cm $\emptyset 
\times$ 64 cm NaI detector and the G\"{o}ttingen recoil detector SENECA in 
coincidence, extracted a backward polarizability value ranging from (-36.5 to 
-39.1)$\times 10^{-4}$ fm$^4$, depending on the parameterization of the 
photomeson amplitudes. These values are consistent with the Mainz measurements 
using the LARA detector~\cite{galler, wolf}, which extracted a backward 
polarizability value ranging from (-37.1 to -40.9)$\times 10^{-4}$ fm$^4$, 
again depending on the parameterization of the photomeson amplitudes. These 
results are also in agreement with the earlier result from Mainz~\cite{olmos}.

The neutron backward spin polarizability was determined to be
\begin{equation}
\ga_{\pi n} = (58.6 \pm 4.0) \times  10^{-4} \, {\rm fm}^4,
\label{eq:gpn}
\end{equation}
from quasi-free Compton scattering on the deuteron~\cite{kossert}. This experiment also used the Mainz 48 cm $\emptyset \times$ 64 cm NaI detector and the 
G\"{o}ttingen recoil detector SENECA in coincidence.

The forward spin polarizability $\ga_0$ is related to the energy-weighted 
integrals of the difference in the helicity-dependent photoreaction 
cross-sections ($\si_{1/2} - \si_{3/2}$) through the following sum rule:
\begin{equation}
\ga_0 = \ga_1 - (\ga_2 + 2\ga_4) =  \frac{1}{4 \pi^2} \int_{\w_{th}}^{\infty} 
\frac{\si_{1/2}-\si_{3/2}}{\w^3}\, \mathrm{d}\w,
\label{eq:g0}
\end{equation}
where $\w_{th}$ is the pion-production threshold. The following results on 
$\ga_0$ were estimated using the VPI-FA93 multipole analysis~\cite{sandorfi}:
\begin{eqnarray}
\ga_{0p} &\simeq& -1.34 \times 10^{-4} \, {\rm fm}^4, 
\label{eq:g0p} \\ 
\ga_{0n} &\simeq& -0.38 \times 10^{-4} \, {\rm fm}^4.
\label{eq:g0n}
\end{eqnarray}

Our goal in this paper is to calculate beam-polarization/deuteron-spin 
dependent observables for the elastic Compton scattering process on the 
deuteron at energies below the pion mass with an emphasis on the role of the 
neutron polarizabilities. In a recent paper, Chen, Ji and Li~\cite{chen}, 
performed calculations in pionless effective theory with a similar goal. 
However, that theory breaks down at $\sim 50$ MeV in contrast to the EFT used 
here which is applicable from around 50 MeV to the pion-production threshold. Knowing that the proton Thomson term dominates the $\ga d$ scattering at low energies (tens of MeVs), it is essential to study the process at energies where the sensitivity to the neutron polarizabilities will be appreciable.

In Section \ref{calc} we describe the calculation with  focus on calculating 
the observables described in Section \ref{obs}. We discuss the strategy used 
for our calculations in Section \ref{strat}. Then in Section \ref{results} we 
report our results pertaining to these observables and also assess some 
theoretical ambiguities. We conclude in Section \ref{conc}.

\section{Our Calculation}
\label{calc}

A comprehensive explanation of the $HB\cpt$ calculation procedure for Compton scattering on the deuteron is available in Ref.~\cite{silas2}. The focus in this section is primarily on the steps that lead to the calculation of photon-polarization/deuteron-spin dependent observables.

\begin{figure}[htbp]
  \epsfig{figure=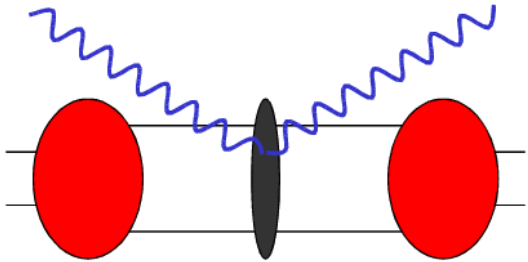,height=4cm}
   \centerline{\parbox{11cm}{\caption{\label{fig1}
The anatomy of the calculation. The irreducible amplitude is computed
in $HB\cpt$ and sewn to external deuteron wavefunctions to give
the matrix element for Compton scattering on the deuteron.
  }}}
\end{figure}

\begin{figure}[htbp]
  \epsfig{figure=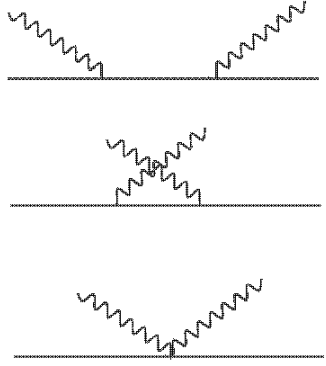,height=5cm}
  \centerline{\parbox{11cm}{\caption{\label{fig2}
Tree level diagrams for the one-body amplitude.
  }}}
\end{figure}

\begin{figure}[htbp]
  \epsfig{figure=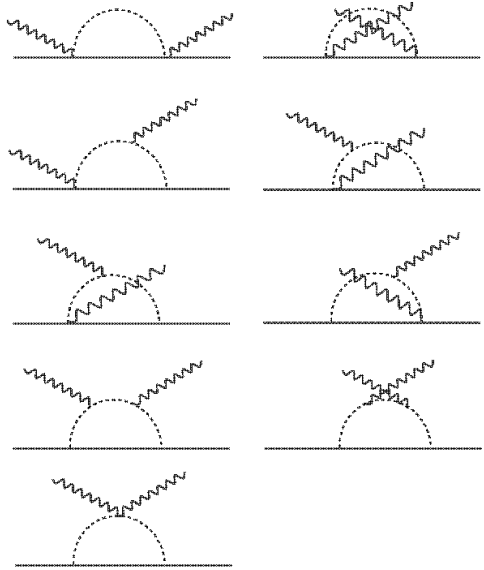,height=8cm}
   \centerline{\parbox{11cm}{\caption{\label{fig3}
Contribution to the one-body amplitude at ${\cal O}(Q^3)$. These diagrams 
contain one pion loop.
  }}}
\end{figure}

\begin{figure}[htbp]
  \epsfig{figure=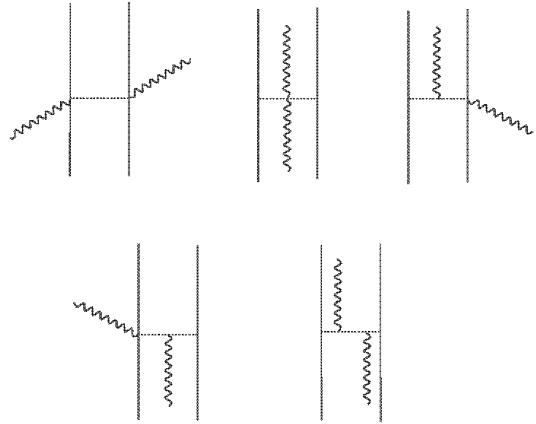,height=6cm}
  \centerline{\parbox{11cm}{\caption{\label{fig4}
Contributing two-body diagrams at ${\cal O}(Q^3)$. Permutations are not shown. 
  }}}
\end{figure}

We proceed by calculating the irreducible amplitude for Compton scattering on 
the deuteron (the central blob in Fig.~\ref{fig1}) and then sew it to the incoming 
and outgoing deuteron wavefunctions to obtain the scattering amplitude, $i.e.$
\begin{equation}
{\cal A} = \bra \Psi'_d|T_{\ga NN}|\Psi_d\ket.
\label{eq:gdamp}
\end{equation}
The irreducible amplitude, $T_{\ga NN}$, consists of one-body interactions 
(see Figs.~\ref{fig2} and \ref{fig3}) and two-body interactions (see 
Fig.~\ref{fig4}),
\begin{equation}
T_{\ga NN} = T_{\ga NN}^{1B} + T_{\ga NN}^{2B}.
\label{eq:gdt}
\end{equation}  
The first term is the one-body amplitude in the scattering process where only one of the nucleons interacts with the photon and the second term is the two-body amplitude where both the nucleons take part in the Compton scattering.

The one-body amplitude in the $\gamma N$ center-of-mass (CM) frame can be 
written as:
\begin{eqnarray}
&&T_{\gamma N}= e^2\left\{ A_1 \vepsprime\cdot\veps 
             +A_2 \vepsprime\cdot\vk \, \veps\cdot\vkp 
             +iA_3 \vsigma\cdot (\vepsprime\times\veps)
             +iA_4 \vsigma\cdot (\vkp\times\vk)\, \vepsprime\cdot\veps
     \right. \nonumber \\
 & & \left.
\!\!\!\!\!\!\!\!\!  
     +iA_5 \vsigma\cdot [(\vepsprime\times\vk)\, \veps\cdot\vkp
            -(\veps\times\vkp)\, \vepsprime\cdot\vk] 
     +iA_6 \vsigma\cdot [(\vepsprime\times\vkp)\, \veps\cdot\vkp
                        -(\veps\times\vk)\, \vepsprime\cdot\vk] \right\}.
\label{eq:Ti}
\end{eqnarray}
where $\veps$($\vepsprime$) and $\vk$($\vkp$) are the unit vectors for 
polarization and momentum of the incoming (outgoing) photons respectively. 
$\vsigma = 2\vec{S}$, where $\vec{S}$ is the quantum-mechanical spin operator 
for the nucleon. The six invariant amplitudes, $A_1 \ldots A_6$, which are 
functions of the photon energy, $\w$, and Mandelstam $t$, are obtained by 
calculating the single-nucleon amplitudes to ${\cal O}(Q^3)$ in $HB\cpt$. The 
relevant Feynman graphs are shown in Figs.~\ref{fig2} and \ref{fig3}.

Defining $\barf =\omega/m_\pi$ and $t=-2{\barf ^2}(1-\cos \theta)$,
where $\theta$ is the center-of-mass angle between the incoming and
outgoing photon momenta, one finds \cite{bkmrev,ulf1,ulf2}:
\begin{eqnarray}
A_1 &=& -\frac{{\cal Z}^2}{M}
        +\frac{g_A^2m_\pi}{8\pi f_\pi^2} 
         \left\{ 1- \sqrt{1-\barf^2}
                 +\frac{2-t}{\sqrt{-t}} 
                  \left[\frac{1}{2} \arctan \frac{\sqrt{-t}}{2}
                        -I_1(\barf,t) \right]\right\},
\nonumber \\
A_2 &=& \frac{{\cal Z}^2 \w}{M^2}
        -\frac{g_A^2 \w^2}{8\pi f_\pi^2 m_\pi} 
         \frac{2-t}{(-t)^{3/2}} 
         \left[I_1(\barf,t)- I_2(\barf,t)\right],
\nonumber \\
A_3 &=& \frac{\omega}{2M^2} 
[{\cal Z}({\cal Z}+2\kappa )-({\cal Z}+\kappa)^2 \cos \theta]
        +\frac{{(2{\cal Z} -1)}g_Am_\pi}{8\pi^2 f_\pi^2} \frac{\barf t}{1-t}
\nonumber \\
    & &  +\frac{g_A^2m_\pi}{8\pi^2 f_\pi^2} 
         \left[ \frac{1}{\barf} \arcsin^2\barf- \barf +2\barf^4 
\sin^2 \theta I_3(\barf,t)\right],
\nonumber \\
A_4 &=& -\frac{({\cal Z}+\kappa )^2\omega}{2M^2}
        +\frac{g_A^2 \w^2}{4\pi^2 f_\pi^2m_\pi} I_4(\barf,t),
\nonumber \\
A_5 &=& \frac{({\cal Z}+\kappa )^2\omega}{2M^2}
        -\frac{{(2{\cal Z} -1)}g_A \w^2}{8\pi^2 f_\pi^2 m_\pi} \frac{\barf}
        {(1-t)}
        -\frac{g_A^2 \w^2}{8\pi^2 f_\pi^2m_\pi}
          [I_5(\barf,t)-2\barf^2\cos\theta I_3(\barf,t)],
\nonumber \\
A_6 &=& -\frac{{\cal Z}({\cal Z}+\kappa )\omega}{2M^2}
        +\frac{{(2{\cal Z} -1)}g_A \w^2}{8\pi^2 f_\pi^2 m_\pi}\frac{\barf}
         {(1-t)}
        +\frac{g_A^2 \w^2}{8\pi^2 f_\pi^2m_\pi}
          [I_5(\barf,t)-2\barf^2 I_3(\barf,t)],
\label{eq:As}
\end{eqnarray} 
where
\begin{eqnarray}
I_1(\barf,t) &=& \int_0^1  dz \,
             \arctan \frac{(1-z)\sqrt{-t}}{2\sqrt{1-\barf^2 z^2}},
\nonumber \\
I_2(\barf,t) &=& \int_0^1  dz \,
             \frac{2(1-z)\sqrt{-t(1-\barf^2z^2)}}{4(1-\barf^2 z^2)-t(1-z)^2},
\nonumber \\
I_3(\barf,t) &=& \int_0^1  dx \, \int_0^1  dz \,
             \frac{x(1-x)z(1-z)^3}{S^3} 
             \left[ \arcsin \frac{\barf z}{R}+ \frac{\barf zS}{R^2}\right],
\nonumber \\
I_4(\barf,t) &=& \int_0^1  dx \, \int_0^1  dz \,
             \frac{z(1-z)}{S}\arcsin \frac{\barf z}{R},
\nonumber \\
I_5(\barf,t) &=& \int_0^1  dx \, \int_0^1  dz \,
             \frac{(1-z)^2}{S}\arcsin \frac{\barf z}{R},
\label{eq:Is}
\end{eqnarray}
with
\begin{equation}
S=\sqrt{1-\barf^2 z^2-t(1-z)^2x(1-x)}, \qquad R=\sqrt{1-t(1-z)^2x(1-x)}.
\label{eq:sr}
\end{equation}
At low energies ($\w < \mpi$), these six invariant amplitudes can be Taylor expanded in $\w$ as follows:
\begin{eqnarray}
A_1 &=& -\frac{{\cal Z}^2}{M}
        +(\alpha + \beta \cos(\theta))\omega^2 + {\cal O}(\w^4),
\nonumber \\
A_2 &=& \frac{{\cal Z}^2\omega}{M^2}
        + \beta \omega^2 + {\cal O}(\w^4),
\nonumber \\
A_3 &=& \frac{\omega}{2M^2} 
[{\cal Z}({\cal Z}+2\kappa )-({\cal Z}+\kappa)^2 \cos \theta] + A_3^{\pi^0} + 
        (\ga_1 - (\ga_2 + 2\ga_4)\cos(\theta))\w^3
        +{\cal O}(\omega^5),
\nonumber \\
A_4 &=& -\frac{({\cal Z}+\kappa )^2\omega}{2M^2} + \ga_2 \w^3 
        +{\cal O}(\omega^5),
\nonumber \\
A_5 &=& \frac{({\cal Z}+\kappa )^2\omega}{2M^2} +  A_5^{\pi^0} + \ga_4 \w^3
        +{\cal O}(\omega^5),
\nonumber \\
A_6 &=& -\frac{{\cal Z}({\cal Z}+\kappa )\omega}{2M^2} +  A_6^{\pi^0} + 
\ga_3 \w^3  +{\cal O}(\omega^5).
\label{eq:Asinw}
\end{eqnarray} 
Here, $A_3^{\pi^0}$, $A_5^{\pi^0}$ and $A_6^{\pi^0}$ are contributions from the $\pi^0$-pole graph, but are not relevant in the current calculations. The zeroth-order term in $\w$ is the leading-order Thomson term (as in Eq.~(\ref{eq:1})) and the first-order term contains contributions from the anomalous magnetic moment. The $\w^2$ term is dependent on $\al$ and $\be$ and the spin polarizabilities ($\ga$'s), appear in the $\w^3$ term. It is fairly evident that, to leading order in $Q$, the nucleon polarizabilities are manifestations of the pion cloud and hence any theory attempting to describe this property of nucleons should have pions as explicit degrees of freedom.  

Since the amplitude in Eq.~(\ref{eq:Ti}) is in the $\gamma N$ CM, it has to be 
boosted to the $\gamma d$ CM frame to obtain $T_{\ga NN}^{1B}$. Consequently, 
\begin{equation}
T_{boost}=-
\frac{{\cal Z}^2 e^2}{2{M^2}\omega}
\lbrack 
\veps\cdot\vkayprime \, \vepsprime\cdot\vkay +
{2}(\veps\cdot{\vec p} \, \vepsprime\cdot\vkay +
\veps\cdot\vkayprime \, \vepsprime\cdot{\vec p})
\rbrack
\label{eq:boost}
\end{equation}
must be added to the one-body amplitude in the $\ga N$ CM frame (Eq.~(\ref{eq:Ti})) when we are computing Compton scattering from the two-nucleon system at ${\cal O}(Q^3)$ in the $\ga d$ CM frame. This additional term is necessary to ensure that the vectors $\veps$ and $\vepsprime$ stay orthogonal to $\vk$ and $\vkp$ respectively in the new frame~\cite{koch}.

The two-body diagrams that contribute to the Compton scattering process at ${\cal O}(Q^3)$ are shown in Fig. \ref{fig4}. The two-body amplitude can be expressed (in the $\gamma d$ CM frame) as:
\begin{equation}
T_{\gamma NN}^{2N}\;=\;-\frac{{e^2}{g_A^2}}{2 f_\pi^2}
\; ({\vec\tau}^{\; 1} \cdot{\vec\tau}^{\; 2}-\tau^{1}_{z}\tau^{2}_{z})
\; ( t^{(a)}+
t^{(b)}+
t^{(c)}+
t^{(d)}+
t^{(e)}),
\label{eq:total}
\end{equation}
where
\begin{eqnarray}
{t^{(a)}}&= &
\frac{{\veps\cdot\vsigone}\;{\vepsprime\cdot\vsigtwo}}
{2\lbrack {\omega^2}-{m_\pi^2}-
(\vpee -\vpeeprime +{\frac{1}{2}(\vkay +\vkayprime )})^2 \rbrack}
+ (1\;\leftrightarrow\; 2), 
\label{ta}\\
{t^{(b)}}&= &  
\frac{{\veps\cdot\vepsprime}\;
\vsigone\cdot ( \vpee -\vpeeprime -{\frac{1}{2}(\vkay -\vkayprime )} )
        \vsigtwo\cdot ( \vpee -\vpeeprime +{\frac{1}{2}(\vkay -\vkayprime )})}
{2\lbrack (\vpee -\vpeeprime -{\frac{1}{2}(\vkay -\vkayprime )})^2 
+{m_\pi^2} \rbrack
\lbrack (\vpee -\vpeeprime +{\frac{1}{2}(\vkay -\vkayprime )})^2  
+{m_\pi^2} \rbrack}
+ (1\;\leftrightarrow\; 2), 
\label{tb}\\
{t^{(c)}}&= &  
-\frac{{\vepsprime\cdot (\vpee -\vpeeprime +{\frac{1}{2}\vkay})}\;
          \vsigone\cdot\veps\;
 \vsigtwo\cdot ( \vpee -\vpeeprime +{\frac{1}{2}(\vkay -\vkayprime )} )}
{\lbrack {\omega^2}-{m_\pi^2}- (\vpee -\vpeeprime 
+{\frac{1}{2}(\vkay +\vkayprime )})^2 \rbrack
\lbrack (\vpee -\vpeeprime +{\frac{1}{2}(\vkay -\vkayprime )})^2  
+{m_\pi^2} \rbrack} \nonumber \\
&+& (1\;\leftrightarrow\; 2), 
\label{tc}\\
{t^{(d)}}&= & 
-\frac{{\veps\cdot (\vpee -\vpeeprime +{\frac{1}{2}\vkayprime} )}\;
\vsigone\cdot ( \vpee -\vpeeprime -{\frac{1}{2}(\vkay -\vkayprime )} )
              \vsigtwo\cdot\vepsprime }
{\lbrack {\omega^2}-{m_\pi^2}- (\vpee -\vpeeprime +
{\frac{1}{2}(\vkay +\vkayprime )})^2 \rbrack
\lbrack (\vpee -\vpeeprime -{\frac{1}{2}(\vkay -\vkayprime )})^2  
+{m_\pi^2} \rbrack} \nonumber \\
&+& (1\;\leftrightarrow\; 2), 
\label{td}\\
{t^{(e)}}&= &  
\frac{2
{\veps\cdot (\vpee -\vpeeprime +{\frac{1}{2}\vkayprime})\;
\vepsprime\cdot (\vpee -\vpeeprime +{\frac{1}{2}\vkay})}\;
\vsigone\cdot ( \vpee -\vpeeprime -{\frac{1}{2}(\vkay -\vkayprime )})\; }
{\lbrack {\omega^2}-{m_\pi^2}- (\vpee -\vpeeprime +
{\frac{1}{2}(\vkay +\vkayprime )})^2 \rbrack
\lbrack (\vpee -\vpeeprime -{\frac{1}{2}(\vkay -\vkayprime )})^2  
+{m_\pi^2} \rbrack}\nonumber \\
&\times& \frac{\vsigtwo\cdot ( \vpee -\vpeeprime +{\frac{1}{2}(\vkay -
\vkayprime )} )}{\lbrack (\vpee -\vpeeprime +{\frac{1}{2}(\vkay -\vkayprime )})
^2  +{m_\pi^2} \rbrack} + (1\;\leftrightarrow\; 2), 
\label{te}
\end{eqnarray}
Here again, $\veps, \vepsprime, \vkay, \vkayprime$ have their usual meaning. $\vpee$($\vpeeprime$) is the initial (final) momentum of the nucleon inside the deuteron, and $\vsigone$($\vsigtwo$) is twice the spin operator of the first (second) nucleon. The numbering of the nucleons is arbitrary and hence ($1\leftrightarrow 2$) denotes the term when the nucleons are interchanged. 

\section{The Observables}
\label{obs}

In this and the subsequent sections we shall use the following convention for the co-ordinate system. The beam direction is defined as the $z$ axis. the $x-z$ plane is the scattering plane with the $y$ axis being the normal to it. For linearly polarized photons, the polarization vectors in the initial state can be along the $x$ or the $y$ axis, $i.e.$, $\veps = \hat{x}$ or $\hat{y}$. For circularly polarized photons in the initial state, the beam helicity, $\la$, can be $\pm1$. The photons are right circularly polarized (RCP) if the beam helicity, $\la = +1$ ($\veps_+ = - \frac{\hat{x}+i\hat{y}}{\sqrt{2}}$) or left circularly polarized (LCP) if the beam helicity, $\la = -1$ ($\veps_- = \frac{\hat{x}-i\hat{y}}{\sqrt{2}}$).

\subsection{Photon/Beam Polarization Asymmetry}

The photon/beam polarization asymmetry uses linearly polarized photons and an unpolarized target. The photon polarization asymmetry is defined as:
\begin{equation}
\Sigma =\frac{(\frac{d\sigma}{d\Omega})_x - (\frac{d\sigma}{d\Omega})_y}{(\frac{d\sigma}{d\Omega})_x + (\frac{d\sigma}{d\Omega})_y},
\label{eq:sigma}
\end{equation}
where the subscript $x$($y$) refers to the differential cross-section when the initial state of polarization of the photon is in (perpendicular) to the plane of scattering. It should be noted that the denominator of Eq.~(\ref{eq:sigma}) is simply the unpolarized total differential cross-section. In calculating $(\frac{d\sigma}{d\Omega})_i$, where $i=(x,y)$ we sum the relevant matrix elements as:
\begin{equation}
\sum_{M', M, \vepsprime = \hat{x},\hat{y}} |\bra M', \vepsprime | T_{\ga NN} |
\veps = \hat{i}, M \ket|^2.
\label{eq:matrixs}
\end{equation}
Here, $M$($M'$) defines the initial (final) spin state of the deuteron and $\vepsprime$ is the final polarization state of the photon.

\subsection{Double-Polarization Asymmetry}

The double-polarization asymmetry involves circularly polarized photons and a spin-polarized target. The expectation is that these observables can provide insight into the spin polarizabilities of the target.

When the target is polarized along the beam direction (parallel or anti-parallel), the corresponding observable is called the parallel target polarization asymmetry and is defined to be:
\begin{equation}
\Sigma_{z,(\la = \pm1)} =\frac{(\frac{d\sigma}{d\Omega})_{\uparrow \uparrow} - 
(\frac{d\sigma}{d\Omega})_{\uparrow \downarrow}}{(\frac{d\sigma}{d\Omega})_
{\uparrow \uparrow} + (\frac{d\sigma}{d\Omega})_{\uparrow \downarrow}}.
\label{eq:sigmaz}
\end{equation}
Parallel (anti-parallel) arrows in the subscript symbolize target polarization parallel (anti-parallel) to the beam helicity. For $\la = \pm 1$ the matrix elements for $(\frac{d\sigma}{d\Omega})_{\uparrow \uparrow}$ are summed as:
\begin{equation}
\sum_{\vepsprime, M'}|\bra M', \vepsprime|T_{\ga NN}|\veps = \hat{\epsilon}_
{\la}, M = +1 \ket|^2,
\label{eq:matrixsz}
\end{equation}
taking the $z$-axis as the spin quantization axis. For $(\frac{d\sigma}{d\Omega})_{\uparrow \downarrow}$, we simply replace $|M=+1\ket$ by $|M=-1\ket$ in the initial state.

The target may be polarized along $\pm \hat{x}$ too. In this case, the observable is called the perpendicular polarization asymmetry and is given by:
\begin{equation}
\Sigma_{x,(\la = \pm1)}=\frac{(\frac{d\sigma}{d\Omega})_{\uparrow \rightarrow} 
- (\frac{d\sigma}{d\Omega})_{\uparrow \leftarrow}}{(\frac{d\sigma}{d\Omega})_
{\uparrow \rightarrow} + (\frac{d\sigma}{d\Omega})_{\uparrow \leftarrow}}.
\label{eq:sigmax}
\end{equation}
Again, the direction of the second arrow in the subscript denotes the target polarization is along the $+\hat{x}$ or $-\hat{x}$ directions. We have to be careful in defining the matrix elements because the spin state of the deuteron is an eigenstate of $S_x$ and not of $S_z$. We express the states $|M_x = \pm 1\ket$ in terms of eigenstates of $S_z$ as:
\begin{equation}
|M_x =\pm 1\ket = \frac{1}{2}[|M_z = +1\ket \pm \sqrt{2}|M_z = 0\ket + 
|M_z = -1\ket].
\label{eq:ketsx}
\end{equation}
Then, as before, for $\la = \pm1$ the matrix elements for $(\frac{d\sigma}{d\Omega})_{\uparrow \rightarrow}$ are summed as: 
\begin{equation}
\sum_{\vepsprime, M'}|\bra M', \vepsprime | T_{\ga NN} | \veps = \hat{\epsilon}_
{\la}, M_x = +1 \ket|^2.
\label{eq:matrixsx}
\end{equation}
For $(\frac{d\sigma}{d\Omega})_{\uparrow \leftarrow}$, we simply replace $|M_x = +1\ket$ by $|M_x = -1\ket$ in the initial state.

It is worth mentioning at this point that in Eqs.~(\ref{eq:sigmaz}) and (\ref{eq:sigmax}) the denominators are not the total unpolarized differential cross-section. This is a manifestation of a spin-1 target. In the case of a spin-$\frac{1}{2}$ target they would represent the total unpolarized differential cross-section. 

\section{Strategy}
\label{strat}

The $\ga d$ elastic scattering process is sensitive only to the isoscalar component of the nucleon polarizabilities. For the purpose of this paper, we shall assume that the proton polarizabilities are established and focus on the neutron polarizabilities. We believe that the debate is not over whether the neutron polarizabilities exist or not, but how accurately they can be extracted. Hence, the goal of this project is to build a strategy to facilitate the accurate extraction of $\al_n$, $\be_n$, and $\ga_{1n} \ldots \ga_{4n}$. The ${\cal O}(Q^3)$ predictions from Eq. (\ref{eq:alphaOQ3}) give a starting point in the search.

An examination of Eq.~(\ref{eq:Asinw}) reveals that the leading-order (${\cal O}(Q^2)$) term is the Thomson term. Since this term is absent for the neutron, its polarizabilities express themselves via interference with the proton Thomson term in coherent $\ga d$ scattering. We know that the unpolarized differential cross-section (dcs) is proportional to the spin-average of the square of the amplitude which makes it  apparent that $\al_n$ and $\be_n$ occur at ${\cal O}(\w^2)$ and the $\ga_n$'s occur at ${\cal O}(Q^4)$ in the dcs. This is also true for $\Si$. Hence we can say that as we increase  the photon energy we should first access $\al_n$, then $\be_n$ (as it is of order $10\%$ of the value of $\al_n$) then finally the $\ga_n$'s in the dcs or in $\Si$. In this paper we focus on $\Si$ only, as much work has already been done on the dcs~\cite{silas1, silas2}. We might expect to be able to extract $\al_n$ and $\be_n$ at intermediate energies ($\sim$100 MeV) and focus on the $\ga_n$'s at the highest possible energies (in our case it is 135 MeV as we want to remain below the pion-production threshold). Since we are limited by the energy range, it is only logical to use $\Si$ to probe the polarizabilities it is most sensitive to, namely $\al_n$ and $\be_n$. However, we shall see that this strategy is not guaranteed to work as cancellations occur in $\Si$. For the $\ga_n$'s we look at other observables, $\Si_z$ and $\Si_x$ as the situation is more promising there. In $\Si_x$ and $\Si_z$, $\al_n$, $\be_n$ and the spin polarizabilities occur at the same order (${\cal O}(\w^3)$). Thus, having foreknowledge of $\al_n$ and $\be_n$, $\Si_z$ and $\Si_x$ seem ideal for the purpose of extracting the $\ga_n$'s. 

We are interested in knowing how accurately the polarizabilities can be extracted and at ${\cal O}(Q^3)$ there are no free parameters. The ${\cal O}(Q^3)$ predictions for the polarizabilities (see Eq.~(\ref{eq:alphaOQ3})) are modified by contributions of higher orders in the chiral expansion~\cite{kumar, ulf2}. To account for the ${\cal O}(Q^4)$ and higher order effects in $\al_n$, $\be_n$, and $\ga_{1n} \ldots \ga_{4n}$, we introduce six new parameters as corrections to the ${\cal O}(Q^3)$ values, which we call $\De \al_n$, $\De \be_n$ and $\De \ga_{in} (i=1,2,3,4)$. For one particular plot only one of these parameters, for instance, $\De \al_n$ is varied with the rest being set arbitrarily to zero. This gives us a measure of the sensitivity of that particular observable to $\De \al_n$. 

\section{Results and Discussion}
\label{results}

\subsection{Photon/Beam Polarization Asymmetry}
\label{ppa}

At ${\cal O}(Q^2)$ the only term that contributes to $\Sigma$ is the Thomson term which has a $\vepsprime \cdot \veps$ structure. Consequently, the photon/beam polarization asymmetry at leading-order is: 
\begin{equation}
\Sigma_0 =\frac{(\cos^2\theta - 1)}{(\cos^2\theta + 1)}.
\label{eq:s0}
\end{equation}
This result is modified by interference of the Thomson term with the next-to-leading order (${\cal O}(Q^3)$) term in the $\ga NN \rightarrow \ga NN$ amplitude. Note that any term in the $\ga NN \rightarrow \ga NN$ amplitude that has a $\vepsprime \cdot \veps$ structure merely alters the strength of the Thomson term in the amplitude and will, on its own, still just lead to the form in Eq.~(\ref{eq:s0}) for the photon beam polarization asymmetry, $\Si$.

\begin{figure}[htbp]
  \vspace{1.5cm}
  \epsfig{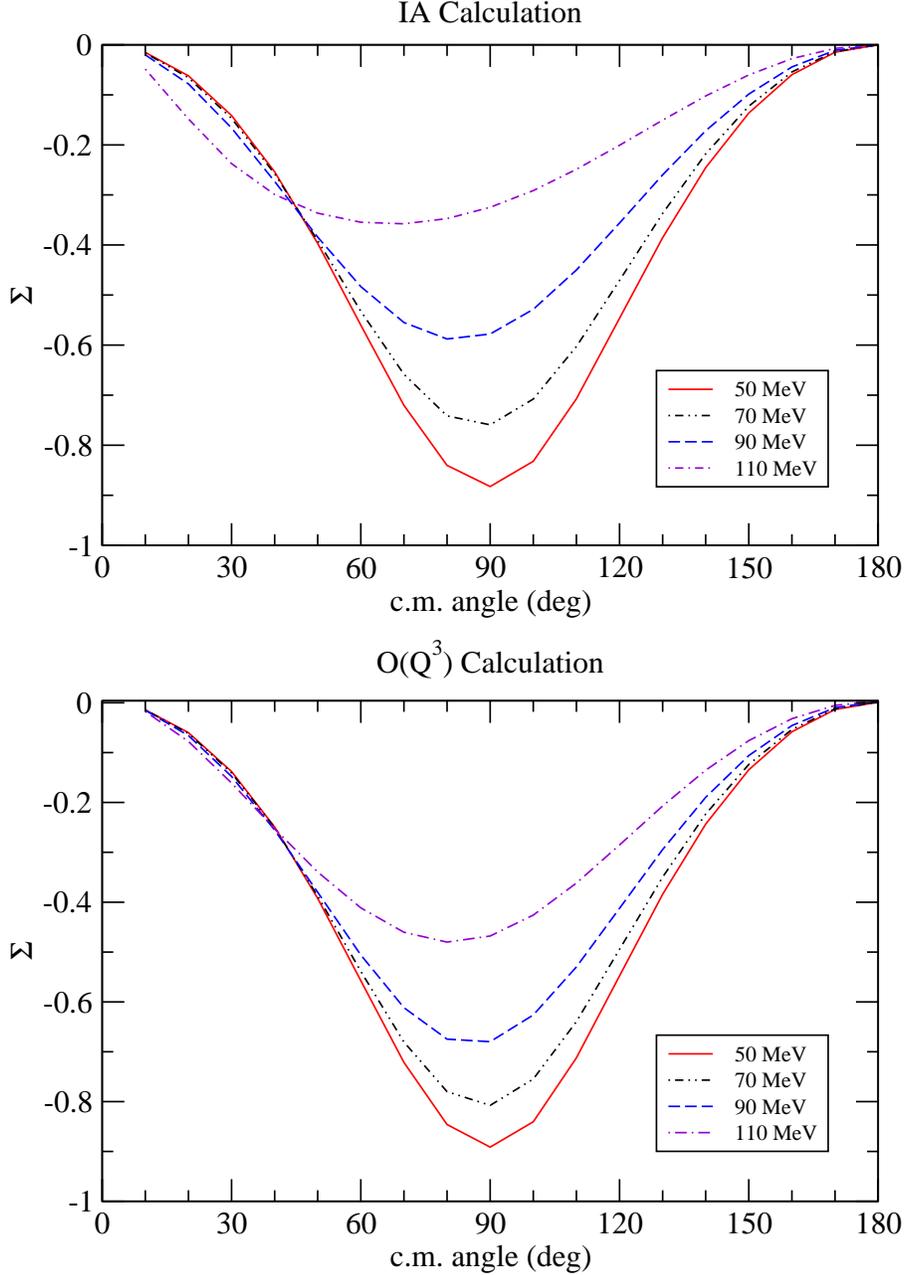}
  \centerline{\parbox{11cm}{\caption{\label{fig5}
Plots of $\Sigma$, in Impulse Approximation and to ${\cal O}(Q^3)$
at different energies.  
  }}}
\end{figure}

Fig.~\ref{fig5} shows the photon polarization asymmetry $\Si$, calculated at various energies in Impulse Approximation and to ${\cal O}(Q^3)$ respectively. The difference between these two plots is that the top one does not have any two-body current in the calculation, $i.e.$, $ T_{\ga NN}^{2B}$ is arbitrarily set to zero. Comparing the two it is fairly obvious that the two-body currents play very little role in this observable up to energies of around 70 MeV. One might wonder why the effects of two-body currents are apparently so small.

As alluded to before, the leading-order result (${\cal O}(Q^2)$ in the chiral expansion) is modified by interference with the next-to-leading order (${\cal O}(Q^3)$). For example, Eq.~(\ref{eq:matrixs}) can be expanded as: 
\begin{equation}
\sum_{M', M, \vepsprime}|\bra M', \vepsprime | T_{\ga NN}^{(2)} | M, \veps = 
\hat{i}\ket|^2 \nonumber \\
+ 2\Re [\sum_{M', M, \vepsprime}\bra M', \vepsprime | T_{\ga NN}^{(2)} |
 M, \veps = \hat{i}\ket\bra M, \veps=\hat{i} | T_{\ga NN}^{(3)} | M', 
\vepsprime \ket] + \ldots,
\label{eq:chexp}
\end{equation}
where the numbers in the superscript  denote the chiral order and $i=(x,y)$. The first term is the Thomson term which produces the result (\ref{eq:s0}) and the second term is the interference between the Thomson term and the next-to-leading order term. Since the Thomson amplitude is non-zero only for $M=M'$, the interference is sensitive to deuteron spin transitions where $\De M=0$. In contrast, the two-body currents dominate the $M = -1 \leftrightarrow M' = 1$ transition of the deuteron ($i.e.$ $\De M=2$). But what is the strength of the two-body currents for the $M=M'$ spin transition amplitudes for the $\ga NN \rightarrow \ga NN$ process? To answer this question we considered the simplest case involving an S-wave deuteron. Focusing on the $\De M=0$ spin transitions, we found that all the three cases ($1 \rightarrow 1,0 \rightarrow 0,-1 \rightarrow -1$) have the same energy behavior and are also roughly of the same strength. $1 \rightarrow 1$ was chosen as the representative because of the relative simplicity in the analytical calculations. To make this particularly simple we also took the limit $\vkay = \vkayprime \rightarrow 0$. In this limit diagrams a, c and d from Fig.~\ref{fig3} do not contribute to the amplitude at all. Only two of the diagrams in Fig.~\ref{fig3} (b and e) contribute and their contributions are:
\begin{eqnarray}
{\cal A}_{(b),(e)}^{xx} &=& \cos \theta \, \xi_{(b),(e)}, \nonumber \\ 
{\cal A}_{(b),(e)}^{xy} &=& 0, \nonumber \\ 
{\cal A}_{(b),(e)}^{yx} &=& 0, \nonumber \\ 
{\cal A}_{(b),(e)}^{yy} &=& \xi_{(b),(e)}.
\label{eq:2b}
\end{eqnarray}
The first (second) letter in the superscript denotes the photon polarization in the initial (final) state. $\xi_{(b),(e)}$ are two different functions of $\vpee, \vpeeprime$ and $\mpi$. The result (\ref{eq:2b}) is obviously of the form $\vepsprime \cdot \veps$. Of course the approximations used to obtain Eq. (\ref{eq:2b}) break down, but numerically we have checked that they are reasonable for $\w \lsim$ 70 MeV. Thus, the two-body currents do not contribute to $\Si$ up to photon energies of around 70 MeV. The fact that the amplitudes in Eq.~(\ref{eq:2b}) have a Thomson-like behavior has been computationally verified where the same assumptions and simplifications were used. Hence, it is safe to conclude that at energies below and up to 70 MeV, an Impulse Approximation calculation does a fair job in predicting the photon/beam polarization asymmetry. In practice, however, we cannot neglect the two-body currents as they may play a significant role in the differential cross-section or when specific spin transitions are considered. Also, our aim is to calculate up to the pion-production threshold and the two-body currents are not negligible over the full energy range.

\begin{figure}[htbp]
  \epsfig{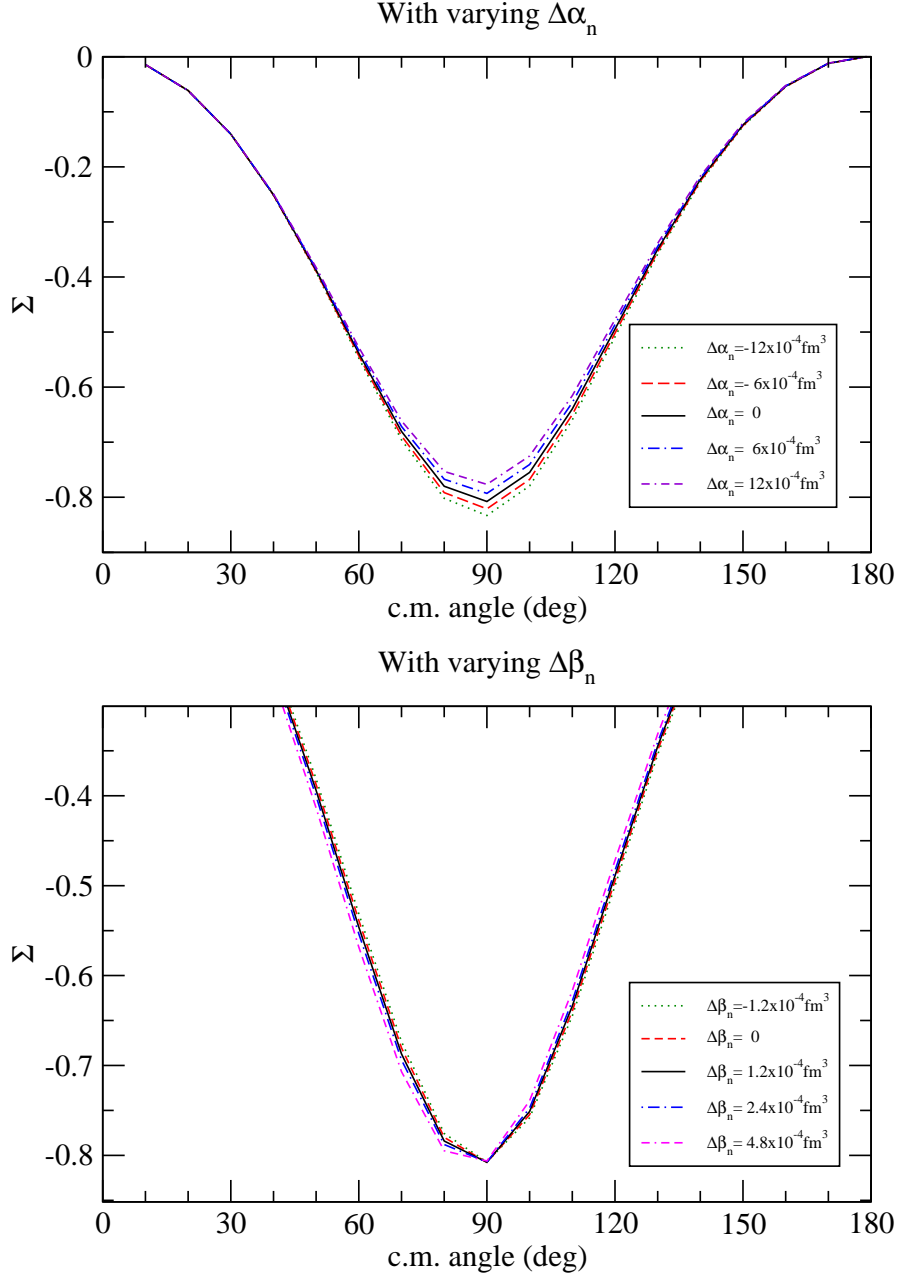}
   \centerline{\parbox{11cm}{\caption{\label{fig6}
Plots of $\Sigma$ for varying $\De \alpha_n$ and $\De 
\beta_n$ at 70 MeV.
 }}}
\end{figure}

Fig.~\ref{fig6} shows the effect of varying $\al_n$ and $\be_n$ on $\Si$. In these plots the values of $\al_n$ and $\be_n$ are varied around their central ${\cal O}(Q^3)$ values, $i.e.$, $\De \al_n$ $(\De \be_n) = 0$ correspond to the value of $\al_n$ $(\be_n)$ given in Eq.~(\ref{eq:alphaOQ3}). Varying the polarizabilities in such a manner gives us an idea as to how accurately these values can be measured. The first observation is that the overall nature of $\Si$ at ${\cal O}(Q^3)$ is not very different from what is obtained at ${\cal O}(Q^2)$, $i.e.$, Eq.~(\ref{eq:s0}). Another obvious observation is that $\Si$ is not appreciably sensitive to variations in either $\al_n$ or $\be_n$. The difference in scales of the top and the bottom plot of Fig.~\ref{fig6} should especially be noted. For $\w \lsim$ 70 MeV we can use Taylor expansions of the amplitudes (as in Eq.~(\ref{eq:Asinw})) and calculate $\Si$ in Impulse Approximation in the c.m. frame. Then by using a Taylor expansion in $\w$ for $\frac{\Si}{\Si_0}$ we can analyze the role of $\al$ and $\be$, the isoscalar nucleon polarizabilities~\cite{reason}. The result is:
\begin{eqnarray}
\frac{\Sigma}{\Sigma_0} &=& 1-\frac{\w}{M}\frac{4\cos\theta}{(\cos^2\theta+1)} 
\nonumber \\
&-& \frac{\w^2}{M^2}\frac{1}{2(\cos^2\theta+1)^2}[8 M^3 \be \cos\theta 
(\cos^2\theta+1) + 2(6 \cos^4\theta - \cos^3\theta - 7\cos^2\theta - 
\cos\theta + 3)
\nonumber \\  
&+& 4 \kp_p(\cos^2\theta+1)(1 - 2 \cos\theta) - \kp_p^2(3\cos^4\theta + 
10\cos^3\theta - 4\cos^2\theta + 10\cos\theta - 7) \nonumber \\
&+& 4\kp_p^3(\cos^2\theta+1)(1-\cos\theta-\cos^2\theta) + 2\kp_n^2
(\cos^2\theta+1)(1-\cos\theta-\cos^2\theta) \nonumber \\
&+& 4\kp_p \kp_n^2(\cos^2\theta+1)(1-\cos\theta-\cos^2\theta) + (\kp_p^2 + 
\kp_n^2)(1- \cos^4\theta)]+ {\cal O}(\omega^3),
\label{eq:taylor}
\end{eqnarray}
where, $\kp_p(\kp_n)$ is the anomalous magnetic moment of the proton (neutron), $M$ is the mass of the nucleon, and $\omega$ is the photon energy. There are three points to be taken away from Eq.~(\ref{eq:taylor})-
\begin{enumerate}
\item $\be$ occurs at ${\cal O}(\w^2)$ but vanishes at $\theta = \frac{\pi}{2}$, as verified by the bottom plot of Fig.~\ref{fig6}. Moreover, according to Eq.~(\ref{eq:alphaOQ3}), $\be$ is approximately one-tenth the value of $\al$ and so its impact on $\Si$ is small, even away from $\frac{\pi}{2}$.
\item $\al$ occurs at the next higher order (${\cal O}(\w^3)$) and hence is suppressed by an additional factor of $\w$. This may seem peculiar since Eq.~(\ref{eq:Asinw}) suggests the occurrence of both $\al$ and $\be$ at the same order. The absence of $\al$ at ${\cal O}(\w^2)$ in Eq.~(\ref{eq:taylor}) is because the terms containing $\al$ in Eq.~(\ref{eq:Asinw}) have an $\vepsprime \cdot \veps$ structure and the $\al$ dependence cancels between the numerator and the denominator in $\Si$ to ${\cal O}(\w^2)$.
\item  Given the second point, one might wonder why the sensitivity to $\al$ seems greater than $\be$. The next term in the Taylor expansion of $\frac{\Sigma}{\Sigma_0}$ contains $\al$ but is multiplied by $\omega/M$. Thus, the ratio of the terms containing $\al$ and $\be$ is proportional to
\begin{equation}
\frac{\al \omega^3/M}{\be \omega^2}=\frac{\al}{\be} \cdot \frac{\omega}{M}
\sim 10 \times 7\% \sim 70\%.
\label{eq:41}
\end{equation}
Thus, given that the range of variation of $\Delta \al_n$ is around four times the range of variation of $\Delta \be_n$, the sensitivity to $\al_n$ appears to be greater.  
\end{enumerate}
The plots in Fig.~\ref{fig6} are a consequence of these facts. The photon/beam polarization asymmetry shows very little sensitivity to $\al_n$ and $\be_n$ at 70 MeV. Hence, the conclusion is that--- at least at this energy---  $\Si$ would not be helpful in extracting the values of $\al_n$ and $\be_n$ with the current levels of experimental precision which is $\sim1\%$ at HI$\gamma$S, especially if we are looking to measure a very small variation in $\alpha_n$ from its ${\cal O}(Q^3)$ predicted value.

\begin{figure}[htbp]
  \epsfig{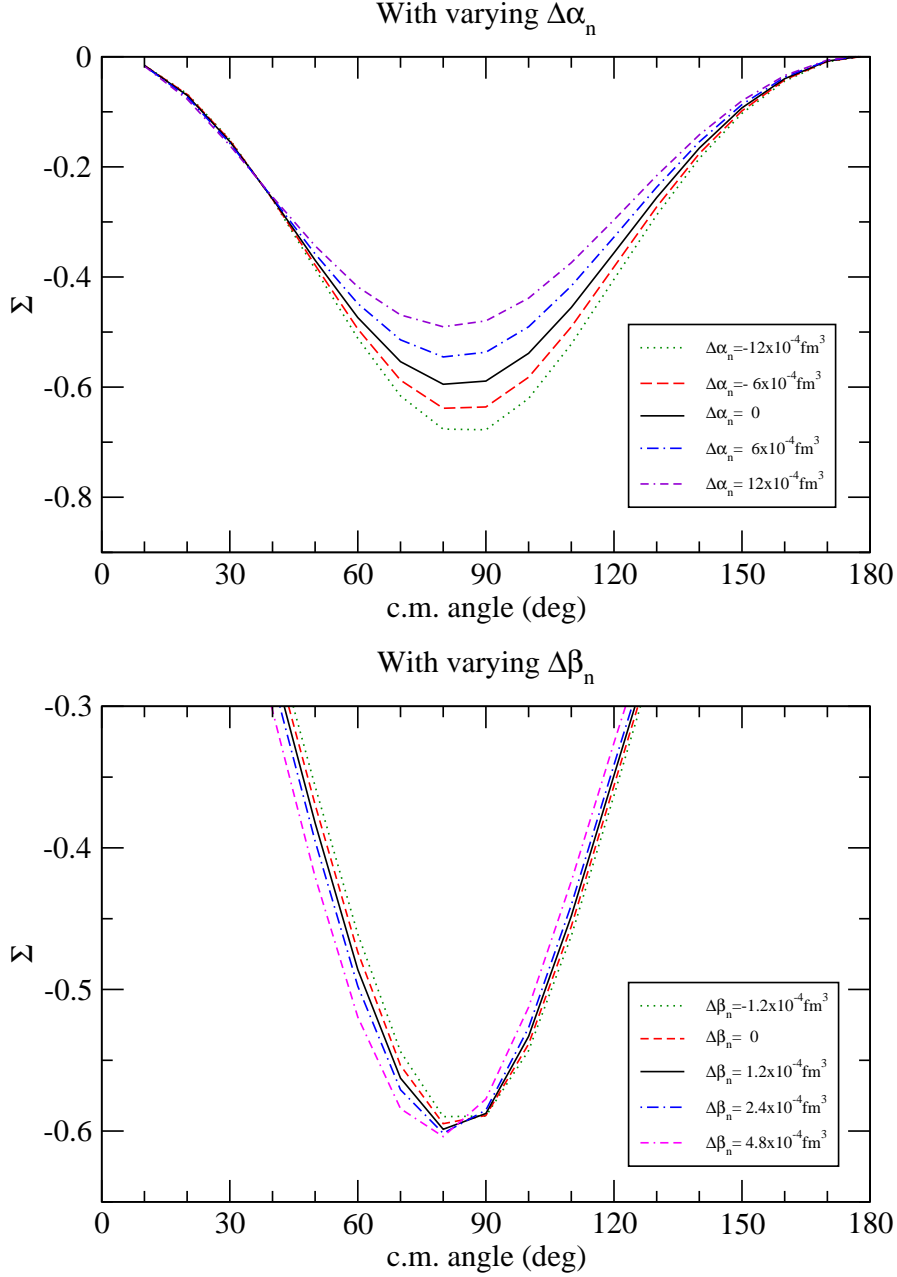}
   \centerline{\parbox{11cm}{\caption{\label{fig6b}
Plots of $\Sigma$ for varying $\De \alpha_n$ and $\De 
\beta_n$ at 100 MeV.
 }}}
\end{figure}

Fig.~\ref{fig6b} shows the same observable $\Sigma$, but at a higher energy (100 MeV), and its sensitivity to $\al_n$ and $\be_n$. As expected, the sensitivity grows with energy. However, the spin polarizabilities start playing a role from $\omega\sim90$ MeV and it is perhaps not advisable to use $\Sigma$ to extract $\al_n$ and $\be_n$ at these energies.  

The sensitivity to the spin-independent isoscalar polarizabilities is actually much stronger in the unpolarized differential cross-section~\cite{silas2}, and so we would advocate using unpolarized Compton scattering on the deuteron to determine $\al_n$ and $\be_n$.

\subsection{Double-Polarization Asymmetry}
\label{dpa}

The spin polarizabilities occur at ${\cal O}(\w^3)$ in the one-body invariant amplitudes. This means that they express themselves in the differential cross-section via products like $\Re(A_i^* A_j), (i,j) = 3\ldots 6$, and consequently they first appear at ${\cal O}(\w^4)$, $i.e.$, at next-to-next-to-next-to-next-to-leading order (N$^4$LO) in $\w$. However, when one considers the double-polarization asymmetries, the leading-order is of ${\cal O}(\w^1)$ and not ${\cal O}(\w^0)$. (Refer to~\cite{bkmrev} for the $\ga p$ case.) Here, the spin polarizabilities express themselves at ${\cal O}(\w^3)$ $i.e.$ NNLO, which means their effects should be seen at much lower energies than in the unpolarized cross-section. It is also true that $\al_n$ and $\be_n$ occur at NNLO in the double-polarization asymmetries and one might expect $\al_n$ to have a greater effect than the spin polarizabilities because it is a few times larger than the $\ga_n$'s. However, the ${\cal O}(\w^3)$ interference terms containing $\al_n$ and $\be_n$ are suppressed by an additional factor of $1/M$ (see Eq.~(\ref{eq:Asinw})) and the effect of variation in these terms is of the same magnitude as from the terms containing the $\ga_n$'s, as long as $\al_n$ and the $\ga_n$'s are varied by the same amount. Thus, the conclusion is that the double-polarization asymmetries are equally sensitive to all of the polarizabilities. This is good news because there are other methods by which $\al_n$ and $\be_n$ can be unambiguously extracted and so at this point we can assume that $\al_n$ and $\be_n$ have been determined and we are interested in extracting the spin polarizabilities alone. It is worth noting here that at low energies ($\w \lsim \mpi/2$) the deuteron observables are dominated by the proton because of the large contribution from the Thomson term. Hence, at these energies, one would expect these observables to look similar to the results for the proton. It is indeed the case that at 70 MeV our results are similar to those of Ref.~\cite{bkmrev} for the proton.

\begin{figure}[htbp]
\vspace{1.5cm}
  \epsfig{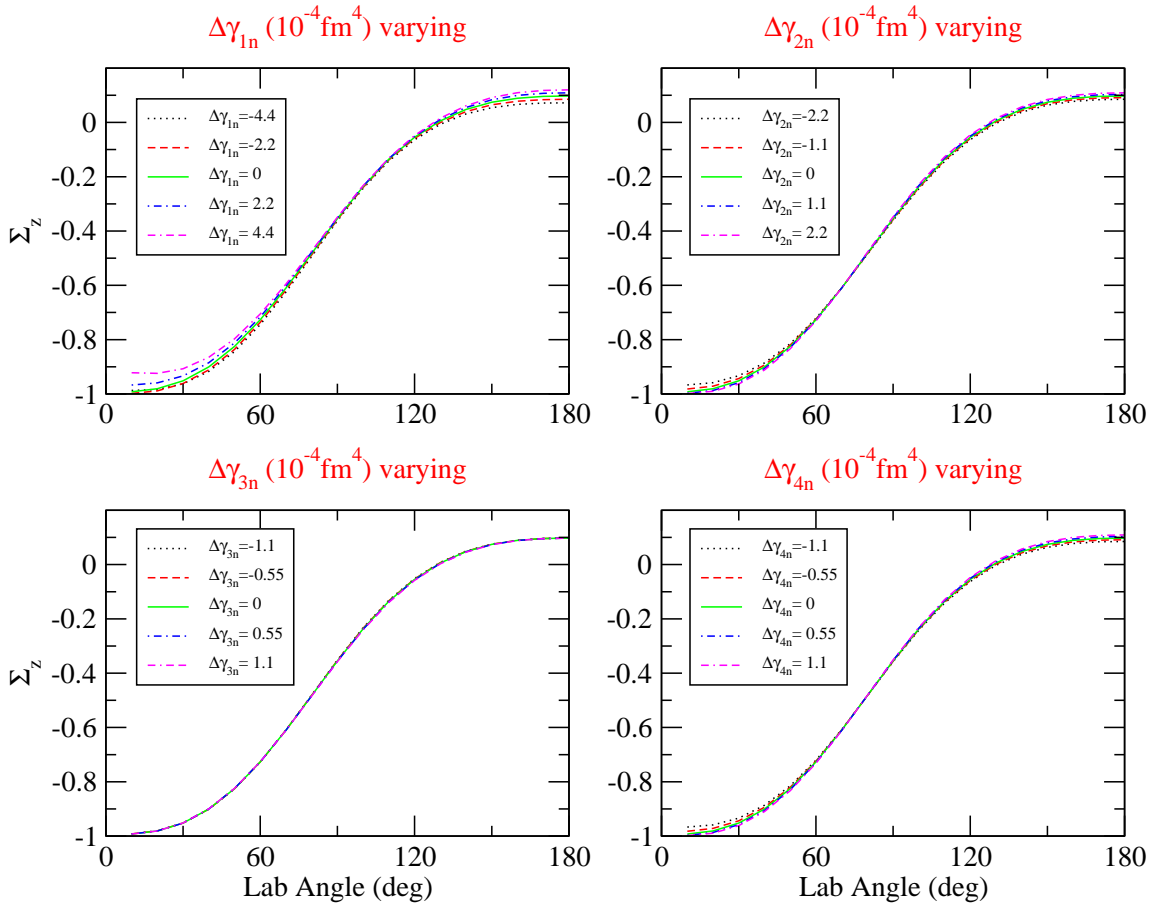}
  \centerline{\parbox{11cm}{\caption{\label{fig7}
Plots of $\Sigma_z$ with varying $\ga_n$'s at 135 MeV (lab) energy.  
  }}}
\end{figure}
\vspace{1cm}
\begin{figure}[htbp]
  \epsfig{figure=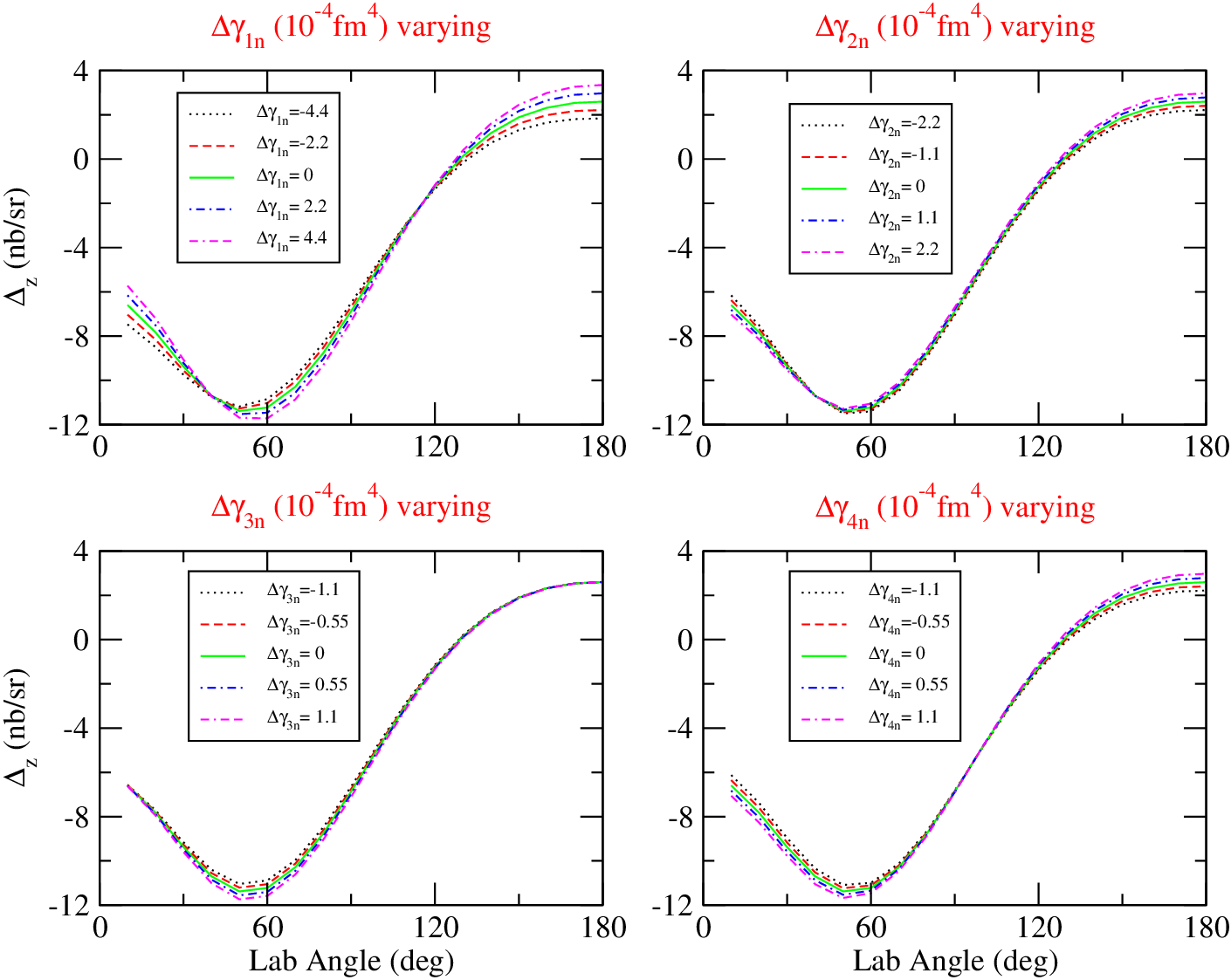,height=12cm}
  \centerline{\parbox{11cm}{\caption{\label{fig8}
Plots of $\De_z$ with varying $\ga_n$'s at 135 MeV (lab) energy.  
  }}}
\end{figure}

The first two sets of plots (Figs.~\ref{fig7} and \ref{fig8}) relate to the parallel target polarization asymmetry. The plots in Fig.~\ref{fig8} only show the numerator of Eq.~(\ref{eq:sigmaz}). It is helpful to look at such a plot because it gives a quantitative estimate of the experimental precision required, but, experimentally speaking, it is better to measure $\Si_z$ as taking the ratio eliminates a number of systematic uncertainties. The top left plot results when $\ga_{1n}$ is varied, the top right corresponds to variation in $\ga_{2n}$, the bottom left corresponds to the variation in $\ga_{3n}$ and the bottom right corresponds to the variation in $\ga_{4n}$. For these and all subsequent plots the values of the neutron spin polarizabilities have been varied around their ${\cal O}(Q^3)$ values, $i.e.$, $\De \ga_{in} = 0$ $(i=1 \ldots 4)$ corresponds to the values in Eq.~(\ref{eq:alphaOQ3}).

Except for $\ga_{1n}$, it seems that there is not an appreciable sensitivity to the neutron spin polarizabilities in $\Si_z$. This is more or less obvious because the sensitivity to the spin polarizabilities is at NNLO in $\w$. The ${\cal O}(Q^3)$ values of the spin polarizabilities predict that $\ga_{1n} > \ga_{2n} > \ga_{3n}(=-\ga_{4n})$, which suggests that the sensitivities will also be in that order, a fact supported by Figs.~\ref{fig7} and \ref{fig8}. Also, since the range of $\De \ga_{in}$ $(i=1 \ldots 4)$ chosen here is $\pm \ga_{in}$, the value of $\ga_{1n}$ was varied the most. However, complications do occur because these observables are sensitive to different combinations of the spin polarizabilities.

\begin{figure}[htbp]
\vspace{0.5cm}
  \epsfig{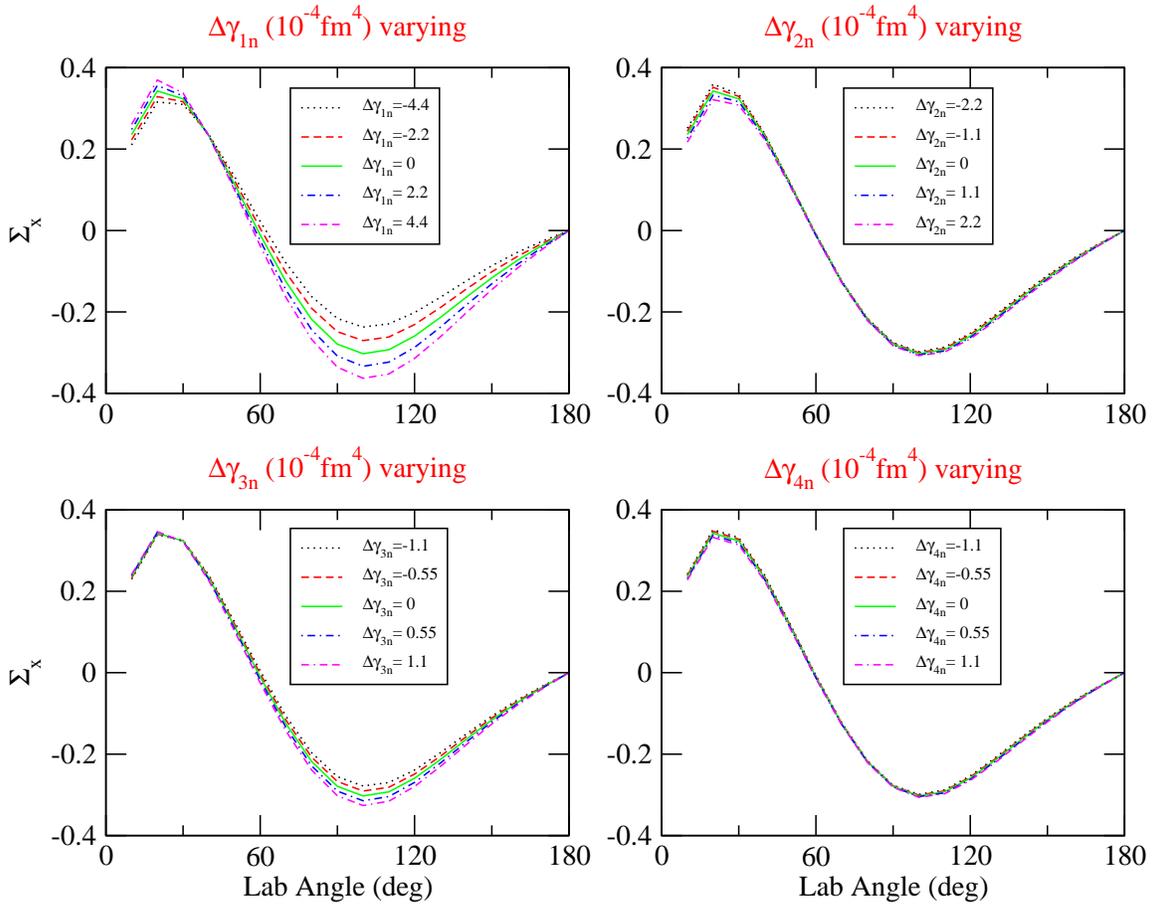}
  \centerline{\parbox{11cm}{\caption{\label{fig9}
Plots of $\Sigma_x$ with varying $\ga_n$'s at 135 MeV (lab) energy.  
  }}}
\end{figure}
\vspace{1cm}
\begin{figure}[htbp]
  \epsfig{figure=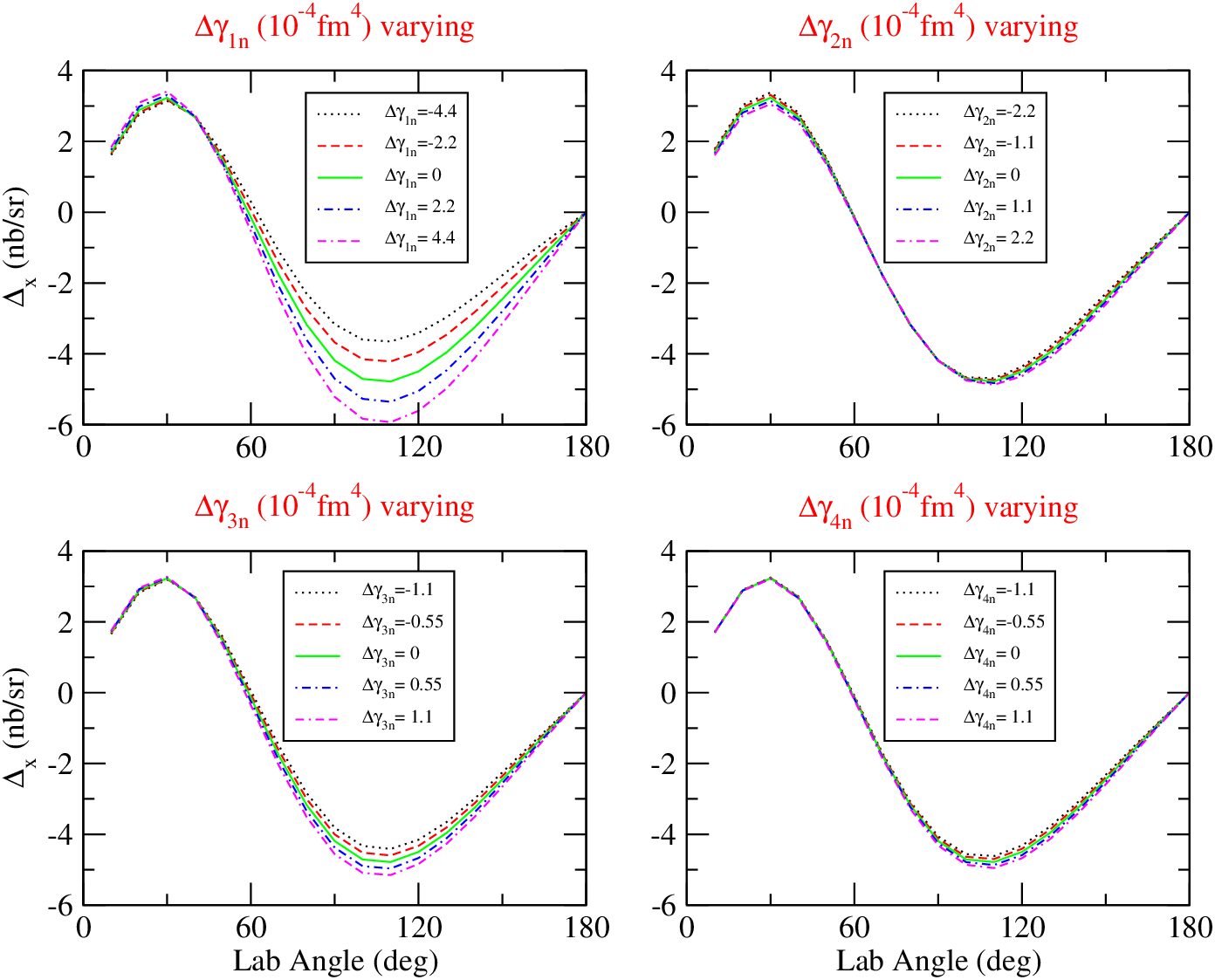,height=12cm}
  \centerline{\parbox{11cm}{\caption{\label{fig10}
Plots of $\De_x$ with varying $\ga_n$'s at 135 MeV (lab) energy.  
  }}}
\end{figure}

Figs.~\ref{fig9} and \ref{fig10} show the perpendicular polarization asymmetry with Fig.~\ref{fig10} showing only the numerator of Eq.~(\ref{eq:sigmax}). Again the neutron spin polarizabilities are varied around their ${\cal O}(Q^3)$ values. As  before, this observable is not very sensitive to variations in the neutron spin polarizabilities, except for $\ga_{1n}$ and $\ga_{3n}$. There is a fair amount of sensitivity to $\ga_{1n}$, but ignorance of $\ga_{3n}$ could hinder extraction of $\ga_{1n}$. The projected HI$\ga$S error-bar for $\Si_x$ with the current flux is $\sim 5\%$. Thus, with the upgraded flux at HI$\ga$S, an experiment to extract some combination $\ga_{1n}$ and $\ga_{3n}$ with good precision from a measurement of $\Si_x$ in elastic $\ga d$ scattering seems feasible. This combined with measurements for $\ga_{0n}$ and $\ga_{\pi n}$ such as mentioned in Eqs.~(\ref{eq:gpn}) and (\ref{eq:g0n}) may make it possible to pin down $\ga_{1n}$, $\ga_{3n}$ and a combination of $\ga_{2n}$ and $\ga_{4n}$.

\subsection{Theoretical Ambiguities}
\label{thamb}

The theoretical tool used in our calculations is $HB\cpt$ in which baryons are introduced with the assumption that their rest four momenta dominate their total four momenta, in other words, the nucleons are virtually at rest. However, in reality this is not true and because of this the pion-production threshold is not in the correct place in the theory. This assumption creates a problem for all processes involving baryons near the pion-production threshold. For instance, the $\ga p$ cross-section varies rapidly near the threshold and if one is to believe the $HB\cpt$ calculations around this energy, it is crucial to get the pion-production threshold in the right place (see, $e.g.$, Ref.~\cite{judith}). The same is true for our calculations too. To remedy this problem Hildebrandt et al.,~\cite{robert2} used
\begin{equation}
\tilde{\w} = \w + \frac{\w^2}{2M},
\label{eq:w1}
\end{equation}
as the energy `going into' the $\ga N$ amplitude as this includes the nucleon kinetic energy which is otherwise neglected in $HB\cpt$.

To determine the appropriate photon energy to use in Eqs.~(\ref{eq:As}), (\ref{eq:Is}) and (\ref{eq:sr}) we examine the integrands in the one-body and the two-body currents and extract the energy at which they first encounter a pole. To do this, firstly we need to define the kinematics. Working in  the $\ga d$ CM frame and defining $k_{\mu} = (\w,\vkay)$ and $k'_{\mu} = (\w',\vkayprime)$ to be the four-momenta of the incoming and the outgoing photon, the four-momenta of the outgoing deuteron will be $K_{\mu} = (-B + \frac{\w^2}{2M_d},-\vkay)$ and $K'_{\mu} = (-B + \frac{\w'^2}{2M_d},-\vkayprime)$, where $B$ is the deuteron binding energy. Next,
\begin{itemize}
\item for the one-body currents, we put the spectator nucleon on its mass shell and calculate the four momenta of the struck nucleon using four-momentum conservation. In the final step, using these momenta, we extract the energy at which the integrand first encounters a pole, which is
\begin{equation}
\w = -B + \mpi + \frac{\mpi^2}{2M_d} + {\cal O}(\mpi^3).
\label{eq:w2}
\end{equation}
\item for the two-body currents, one of the nucleons is put on its mass shell and the four-momenta of the nucleon and of the pion(s) are calculated using four-momentum conservation. As in the one-body case, the energy at which the integrand encounters a pole is extracted. Interestingly, the energy turns out to be the same as in Eq.~(\ref{eq:w2}).
\end{itemize}
Hence, for the purpose of our calculations, we redefine
\begin{equation}
\tilde{\w} = -B + \w + \frac{\w^2}{2M_d}
\label{eq:wtilde}
\end{equation}
as the energy `going into' the $\ga N$ amplitude. In all of the results reported in the previous two subsections, the value of $\w$ was redefined according to Eq.~(\ref{eq:wtilde}).

\begin{figure}[htbp]
\vspace{1.0cm}
  \epsfig{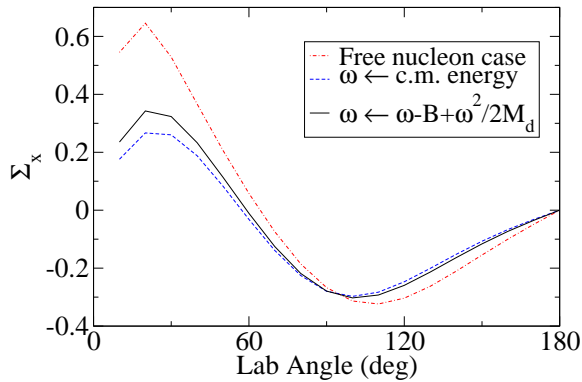}
  \centerline{\parbox{11cm}{\caption{\label{fig11}
Plots of $\Sigma_x$ at 135 MeV (lab) energy. The dashed curve is the $HB\cpt$ calculation where $\tilde{\w}$ is simply the photon energy in the $\ga d$ c.m. frame. The dot-dashed curve symbolizes the other extreme where the nucleons are assumed to be free. The solid curve represents the calculation done with $\tilde{\w}$ redefined as in Eq.~(\ref{eq:wtilde}).  
  }}}
\end{figure}

Fig.~\ref{fig11} is a representation of how observables are affected by altering the value of the CM energy. The dashed curve is the $HB\cpt$ calculation where the nucleon is assumed to be heavy and hence practically at rest, $i.e.$ $\tilde{\w} = \w$. The dot-dashed curve symbolizes the other extreme where the nucleons are assumed to be free and so $\tilde{\w} = \w + \frac{\w^2}{2M}$. The real world  (for nucleons bound inside a deuteron nucleus) is somewhere in between and the solid curve represents such a situation. The uncertainty arising from the freedom to adopt different prescriptions for $\tilde{\w}$ is an effect of order ${\cal O}(Q^4)$ but it obviously has a significant effect on our conclusions. The impact of this ambiguity is energy-dependent and largest at 135 MeV. For a realistic estimate of this uncertainty, one should consider  the discrepancy between the solid and the dashed curve.

\begin{figure}[htbp]
\vspace{1.0cm}
  \epsfig{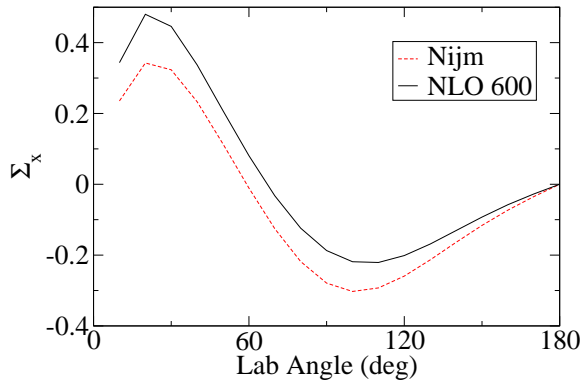}
  \centerline{\parbox{11cm}{\caption{\label{fig12}
This figure shows $\Sigma_x$ at 135 MeV (lab) energy for two different wavefunctions.  
  }}}
\end{figure}

Based on $NN$ potentials that describe the low-energy $NN$ data quite accurately, there are several deuteron wavefunctions on the market. All of these wavefunctions differ from one another in facets like the deuteron D-state probability. Because of these inherent differences, the results reported in this paper will be sensitive to the choice of deuteron wavefunctions. This too is a higher order effect and should typically enter the result for the $\ga d$ process only at ${\cal O}(Q^5)$. We have used the wavefunctions obtained from the Nijm93 potential model, which gives a very good description of the $np$ data to energies of 350 MeV with a $\chi^2$ per datum of 1.87~\cite{stoks}. Fig.~\ref{fig12} gives a flavor of the effect that choosing a different wavefunction has on a representative observable. The solid curve uses the next-to-leading order chiral wavefunction~\cite{ulf3}, whereas the dashed curve uses the Nijmegen wavefunction. It was shown in Ref.~\cite{silas2} that these two wavefunctions provide the extrema between which the uncertainty due to the choice of wavefunctions lie. It should be noted that the uncertainty is largest for the case presented here, $i.e.$, for 135 MeV.

Comparing Figs.~\ref{fig9} and \ref{fig12} one might be led to the conclusion that though the theory predicts some sensitivity to $\ga_{1n}$, the combined theoretical ambiguities may nullify the effect. However, it is fair to say that the errors presented above depict the largest uncertainties possible and hence give a very conservative estimate of the typical size of the theoretical uncertainties. A reasonable range of the error for $\Si_x$ in the area where there is maximum sensitivity to $\ga_{1n}$ is $\sim 13.8 \%$ from the uncertainty because of the choice of wavefunction and $\sim 1.8\%$ due to the ambiguity in the position of the pion-production threshold in $HB\cpt$. Combining these two uncertainties in $\Si_x$ translates into $\sim 50\%$ uncertainty in the extraction in the value of $\ga_{1n}$. However, much thought is being given toward resolving these issues and hence with the passage of time there can only be reduction in the sizes of the theoretical ambiguities discussed in this Section.

\section{Conclusion}
\label{conc}

From the discussion in the first part of Section~\ref{results} we conclude that the photon/beam polarization asymmetry is not a good observable to extract $\al_n$ and $\be_n$. This observable is not sensitive to $\al_n$ and $\be_n$ because effects of $\al_n$ cancel in this observable at ${\cal O}(\w^2)$ and sensitivity to $\be_n$ disappears at $\theta = \frac{\pi}{2}$ where $\Si$ is largest. However, the second part of Section~\ref{results} leads us to believe that the double-polarization observables, especially $\Si_x$, can be instrumental in extracting information about the neutron polarizabilities. There is real hope in this direction, as the HI$\ga$S upgrade program will improve the experimental precision, while focus on resolving the theoretical ambiguities will reduce the uncertainties discussed in Section~\ref{thamb}.

However, it is evident that the photon polarization asymmetry and double-polarization asymmetries are not sufficient by themselves to extract all the neutron polarizabilities. We advocate first trying to extract $\al_n$ from measurements of the isoscalar polarizabilities from unpolarized cross-section measurements on the deuteron. Then the dispersion sum rule (\ref{eq:baldinn}) can be used to extract $\be_n$. This is not an entirely desirable method because the sum rule for the neutron has itself been derived from the deuteron information and contains model assumptions. But, it should provide a starting point.

With the spin polarizabilities the case is worse because the proton spin polarizabilities are yet to be established. However, plans to measure spin-dependent polarizabilities using $\vec{\ga}\vpee$ scattering are in the pipeline~\cite{higs}. Hopefully, soon spin polarizability measurements for the proton produce promising data. Even with the values for $\ga_{1p} \ldots \ga_{4p}$ in hand, measurements of $\vec{\ga}\vec{d} \rightarrow \ga d$ are not guaranteed to give accurate results, although we have found evidence which suggests there will be sensitivity to neutron spin polarizabilities, especially $\ga_{1n}$, in the $\vec{\ga}\vec{d} \rightarrow \ga d$ data. There is also light at the end of the tunnel in the form of `effective neutron targets' like polarized Helium-3, or the use of breakup reactions~\cite{robert1} involving the deuteron or Helium-3. Though cross-sections in the breakup reactions are significantly smaller than in the elastic case, an enhanced flux at HI$\ga$S will surely help the matter.

To summarize, the long-term plan to extract the neutron polarizabilities should involve
\begin{itemize}
\item extracting $\al_n$ and $\be_n$ from unpolarized experiments.
\item measuring $\ga_{ip}$ $(i=1 \ldots 4)$ for the proton.
\item pinning down a linear combination of $\ga_{1n}$ and $\ga_{3n}$ from $\Si_x$ measurements on deuteron.
\item further work on $\ga_0$ and $\ga_{\pi}$ for both proton and neutron.
\end{itemize}

\bigskip

\section*{Acknowledgments}
We thank H.~Gao, M.~C. Birse, J.~A. McGovern and J.~Feldman for useful discussions relating to the work presented in this paper. We also thank E.~Epelbaum for the chiral NLO wavefunctions and V.~Stoks for the Nijmegen wavefunctions. DC would like to thank TUNL and Duke University for hospitality and an opportunity to understand the experimental intricacies which would have been difficult otherwise and V.~Pascalutsa for his help in the initial stages of this work. This work was carried out under grants DE-FG02-93ER40756 and DE-FG02-02ER41218 of the US-DOE.


\begin{thebibliography}{99}
\bibitem{reason}
We used the isoscalar nucleon polarizabilities here to simplify the expression.
\end{thebibliography}
\end{document}